\documentclass[prd,reprint, nofootinbib,amsmath,amssymb, aps, floatfix]{revtex4-1}
\usepackage{amsmath}
\usepackage{natbib}
\usepackage{graphicx}
\usepackage{amssymb}
\begin{document}
\title{Quantum Penrose Inequality}
\author{Raphael Bousso and Arvin Shahbazi-Moghaddam}
% \email{Correspondence: bousso@lbl.gov}
\affiliation{Department of Physics, University of California, Berkeley, CA 94720, USA}
\affiliation{Lawrence Berkeley National Laboratory, Berkeley, CA 94720, USA}
\author{Marija Toma\v{s}evi\'{c}}
\affiliation{Departament de F\'isica Qu\`antica i Astrof\'isica and Institut de Ci\`encies del Cosmos (ICCUB), Universitat de Barcelona, Mart\'i i Franqu\`es 1, E-08028 Barcelona, Spain}

% \author[a,b]{Arvin Shahbazi-Moghaddam,}

% \author[c]{and Marija Toma\v{s}evi\'{c}}
% \affiliation[c]{Departament de F\'isica Qu\`antica i Astrof\'isica and Institut de Ci\`encies del Cosmos (ICCUB),\\
% Universitat de Barcelona, Mart\'i i Franqu\`es 1, E-08028 Barcelona, Spain}

\begin{abstract}
The classical Penrose inequality specifies a lower bound on the total mass in terms of the area of certain trapped surfaces. This fails at the semiclassical level. We conjecture a Quantum Penrose Inequality: the mass at spatial infinity is lower-bounded by a function of the generalized entropy of the lightsheet of appropriate quantum trapped surfaces. This is the first relation between quantum information in quantum gravity, and the total energy.
\end{abstract}
\maketitle

\section{Introduction}
\label{sec-intro}

Many theorems in classical General Relativity involve the area or expansion of codimension 2 surfaces. This includes Hawking's area theorem~\cite{Haw71}, the focussing theorem~\cite{Wald}, certain singularity theorems~\cite{Pen64}, and the area theorem for holographic screens~\cite{BouEng15a,BouEng15b}. The classical theorems rely on the null energy condition (NEC), that $T_{ab} k^a k^b\geq 0$ at every point in the spacetime, where $T_{ab}$ is the stress tensor of matter and $k^a$ is any null vector.

In this paper we will study the Penrose inequality~\cite{PenNS,Mar09}, a generalization of the positive mass theorem~\cite{SchYau81} in asymptotically flat spacetimes. The Penrose inequality relates the area of any (suitably minimal~\cite{EngHor19,BigQPI}) marginally trapped surface $\mu$ to the total mass defined at spatial infinity~\cite{ADM}.
\begin{equation}
  m\geq \sqrt{\frac{A[\mu]}{16\pi G^2}}~.
  \label{eq-piintro}
\end{equation}
This conjecture was originally motivated as a method for testing the cosmic censorship conjecture. It has not been proven except in special cases~\cite{BraSch00}. Here we consider it in its own right, as a quantitative relation between geometry and energy.

Quantum fields are known to violate the NEC in bounded regions. Indeed, the above theorems are all known to fail in the presence of quantum matter. However, in each case there exist semiclassical generalizations that appear to remain valid: respectively, the Generalized Second Law (GSL) for event horizons~\cite{Bek72,Bek73,Bek74}, the Quantum Focussing Conjecture~\cite{BouFis15a}, Wall's Quantum Singularity Theorem~\cite{Wal13}, and the GSL for Q-screens~\cite{BouEng15c}. Remarkably, limits of these semiclassical conjectures can yield nontrivial, provable new results in relativistic quantum field theory, such as the Quantum Null Energy Condition~\cite{BouFis15b,KoeLei15,BalFau17,CeyFau18}. 

Like the positive mass theorem, the Penrose inequality assumes that matter has positive energy density. We will demonstrate explicitly that Eq.~(\ref{eq-piintro}) can be violated by quantum matter. The violation can be large: the mass at infinity can be less than the lower bound set by the area, by a relative fraction that is $O(1)$, not $O(\hbar)$. Thus we show that the classical Penrose inequality does not hold true in Nature.

We will then propose a quantum-corrected Penrose inequality, in which the area is replaced by the generalized entropy on the lightsheet~\cite{CEB1} of an appropriate surface. We will provide preliminary evidence for its validity. A separate article~\cite{BigQPI} provides more background and technical details, presents further tests, discusses the relation to the cosmic censorship conjecture, and considers the analogue of the Penrose inequality in asymptotically Anti-de Sitter space (see also~\cite{EngHor19}).

The generalized entropy is expected to have a simple representation in terms of the fundamental degrees of freedom in quantum gravity~\cite{RyuTak06,EngWal14,EngWal17,BouSha19}. Our proposal thus elevates Eq.~(\ref{eq-piintro}) to a relation between information and energy. This breaks new ground: known relations between information and energy~\cite{Bek81,Bou03,Cas08,BouFis15b} do not involve Newton's constant. There are also relations between information and spacetime geometry~\cite{Tho93,Sus95,FisSus98,CEB1,CEB2} but those do not involve the energy.

\section{Violating the Penrose Inequality With Quantum Matter}

\begin{figure}%
%     \centering
     \includegraphics[width=.35\textwidth]{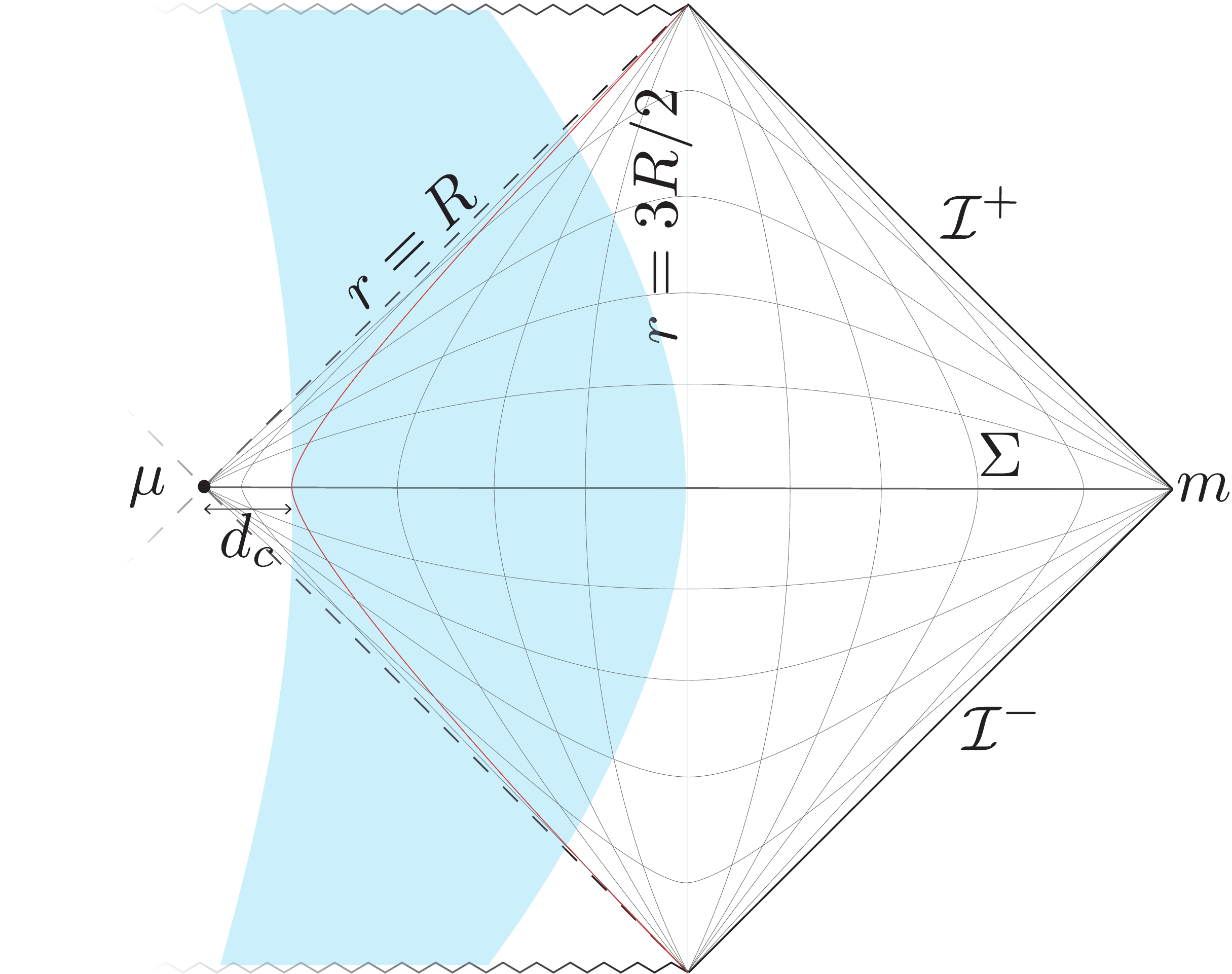}
     \caption{Penrose diagram of the Schwarzschild solution. The Boulware state has negative energy (blue) near the black hole, outside of the cutoff sphere $d_c$ at $t=0$. This decreases the mass $m$ at infinity by an $O(1)$ fraction of the classical black hole mass $M$, violating the classical Penrose inequality.}%
     \label{fig-flatBoulware}%
\end{figure}

To demonstrate a violation of Eq.~(\ref{eq-piintro}) by quantum effects, we consider the maximally extended Schwarzschild black hole with mass $M=R/2G$. Let $\mu$ be the bifurcation surface where the two event horizons intersect. See Fig.~\ref{fig-flatBoulware}. We will consider a perturbative state, whose contribution to the mass $m$ can be computed by integration over a Cauchy surface $\Sigma$ of the exterior~\cite{Wald}:
\begin{equation}
  \Delta m = \int_\Sigma d^3\Sigma \, T_{ab}n^a\xi^b~.
  \label{eq-genint}
\end{equation}
Here $d^3\Sigma$ is the volume element on $\Sigma$, $n^a$ is the unit normal to $\Sigma$, and $\xi^b$ is the timelike Killing vector field $\partial_t$. The classical Penrose inequality is saturated, $m=M=(A[\mu]/16\pi G^2)^{1/2}$, if the stress tensor vanishes everywhere. Thus, any net negative energy of the quantum fields, $\Delta m<0$, will lead to a violation of Eq.~(\ref{eq-piintro}).

For a two-sided black hole, one often considers the Hartle-Hawking state, in which thermal radiation is present both in the past and the future. In the Unruh state, no radiation is sent in. Here we consider the Boulware vacuum, in which neither ingoing nor outgoing radiation is present far from the black hole. Then the energy density near the horizon is negative for a massless scalar field~\cite{Boulware,Candelas}: $T_{tt} \sim -\hbar R^{-4}(1-R/r)^{-1}$.
%  \label{eq-boulstress}

We regulate the divergence at $r=R$ by cutting off the Boulware state at a proper distance $d_c\ll R$ outside of $\mu$, at $r=R+d_c^2/4R$. For full control of the semiclassical expansion, we choose $d_c\gg l_P$. Physically, this implies that the state will behave like the Boulware state only for a limited time of order $R\log (R/d_c)$. Inside of the cutoff sphere, we take the state to be the Hartle-Hawking state, with vanishing stress tensor. There will be some corrections at the cutoff sphere, but the two regions can be glued without introducing large backreaction or compensating positive contributions to the mass at infinity~\cite{BigQPI}.

Choosing $\Sigma$ to be a surface of constant $t$, and neglecting factors of order unity, one finds
\begin{equation}
  \Delta m \sim - R^2 \int_{R+d_c^2/4R}^{3R/2} dr \frac{\hbar}{R^4} \left(1-\frac{R}{r}\right)^{-2} \sim -\alpha M~,
\end{equation}
where $\alpha = l_P^2/d_c^2$.
Thus, one can reduce the mass at infinity from $M$ to $(1-\alpha) M$, violating Eq.~(\ref{eq-piintro}). The violation is parametrically small, $\alpha\ll 1$, but substantial in that it is not $O(\hbar)$ but $O(1)$. The contribution from each mode is $O(\hbar)$; but the number of available modes in the thermal atmosphere, at fixed control parameter $l_P/d_c\ll 1$, is $O(\hbar^{-1})$. Thus, the negative energy of the quantum fields can cancel off an $O(1)$ fraction of the black hole's classical mass.

\section{Quantum Penrose Inequality}
\label{sec-QPI}

\subsection{Generalized Entropy and Quantum Expansion}
\label{sec-qnd}

We now introduce quantum generalizations of certain geometric quantities, necessary for formulating a Quantum Penrose Inequality. More details can be found in~\cite{BouFis15a}. 

The key ingredient is the notion of \textit{generalized entropy} $S_{\text{gen}}$, first introduced~\cite{Bek72,Bek73} as the total entropy of a black hole and its exterior on a given time slice. One can extend its definition to apply not only to the horizon of a black hole, but to any Cauchy-splitting surface $\sigma$:
\begin{equation}
  S_{\text{gen}} \equiv \frac{A[\sigma]}{4G\hbar} + S_{\text{out}} + \dots~,
  \label{eq-gedef}
\end{equation}
where $A[\sigma]$ is the area of $\sigma$, and
$ S_{\text{out}}= - {\rm Tr}\, \rho_{\rm out} \log  \rho_{\rm out} $
is the von Neumann entropy of the state of the quantum fields, restricted to one side of $\sigma$: $\rho_{\rm out} = {\rm Tr}_{\overline{\rm out}}\, \rho$.
Here $\rho$ is the global quantum state, and the trace is over the complement region, $\overline{\rm out}$.

The leading divergence in $S_{\rm out}$ is given by $A/\epsilon^2$, where $\epsilon$ is a short-distance cutoff. It comes from entanglement in the vacuum across $\sigma$~\cite{Sor83,Sre93}. The geometric term in Eq.~(\ref{eq-gedef}) can be thought of as a counterterm. The dots indicate the presence of subleading divergences in $ S_{\rm out}$ which come with their own geometric counterterms,
% . One expects that the divergences coming from the renormalization of $G$ and from short-distance entanglement cancel out~\cite{BouFis15a},
so that $S_{\text{gen}}$ is a finite and well-defined quantity~\cite{SusUgl94,BarFro94}. 

In a semiclassical expansion in $G\hbar$, Eq.~(\ref{eq-gedef}) defines a quantum-corrected area of the surface $\sigma$:
\begin{equation}
  A_Q[\sigma] \equiv 4G\hbar S_{\rm gen} = A[\sigma]+4G\hbar S_{\text{out}} + \dots~.
\end{equation}
Thus, one can use the generalized entropy to incorporate quantum effects into various geometrical objects that derive from the area of surfaces. For example, the classical expansion~\cite{Wald} of a surface $\sigma$ at a point $y \in \sigma$ can be defined as a functional derivative of the area~\cite{BouFis15a},
\begin{equation}
    \theta[\sigma; y] = \frac{1}{\sqrt{h(y)}}\frac{\delta A[V]}{\delta V(y)}~,
\end{equation}
where $h$ is the area element of the metric restricted to $\sigma$. The function $V(y)$ specifies the affine location of $\sigma$ and nearby surfaces along a congruence of null geodesics orthogonal to $\sigma$.
% The derivative is taken with respect to a function $V(y)$ that defines a surface that lies an affine parameter distance $\lambda = V$ away from $\sigma$, along the null geodesic which emanates from $\sigma$ at $y$. \par
The above definition unnecessarily involves the entire surface $\sigma$, though $\theta$ depends only on the local curvature at $y$. However, by substituting $A\to A_Q$, we can now define the {\em quantum expansion}, $\Theta$, which does depend on all of $\sigma$:
\begin{equation}
    \Theta[\sigma; y] \equiv \frac{4G\hbar}{\sqrt{h(y)}}\frac{\delta S_{\text{gen}}[V]}{\delta V(y)}~.
\end{equation}
  
Let $\Theta_\pm$ be the quantum expansion of the future-directed light-rays orthogonal to a surface $\mu_Q$, to either side. If $\Theta_+ \leq 0$ ($\Theta_+ = 0$) and $\Theta_- \leq 0$, then we call $\mu_Q$ a \textit{(marginally) quantum trapped surface}. This is analogous to the definition of trapped surfaces using $\theta$.

Quantum trapped surfaces, in the semiclassical setting, enjoy some of the properties obeyed by trapped surfaces in the classical setting. For example, trapped surfaces cannot lie outside the black hole. When the NEC is violated, they can; but quantum trapped surfaces must still lie inside or on the horizon~\cite{Wal13}. This will be important for our formulation of the quantum Penrose inequality.

\subsection{Formulation}
\label{sec-formulation}

We will now obtain a Quantum Penrose Inequality (QPI), in three steps. First, we replace the area with generalized entropy in Eq.~(\ref{eq-piintro}): $A \to A_Q = 4G\hbar\, S_{\rm gen}$. Secondly, we specify the surfaces to which the inequality can be
applied. Instead of a marginally trapped surface, in the QPI it is natural to consider any surface $\mu_Q$ that is quantum marginally trapped (subject to an appropriate generalization of the minimality condition that also applies in the classical case~\cite{BigQPI}). 

%We also demand that the surface $\mu_Q$ minimize the generalized entropy on some Cauchy surface of its outer wedge $O_W[\mu_Q]$.

\begin{figure}%
%\centering
%   \subfloat[label 1]{{
    \includegraphics[width=.3\textwidth]{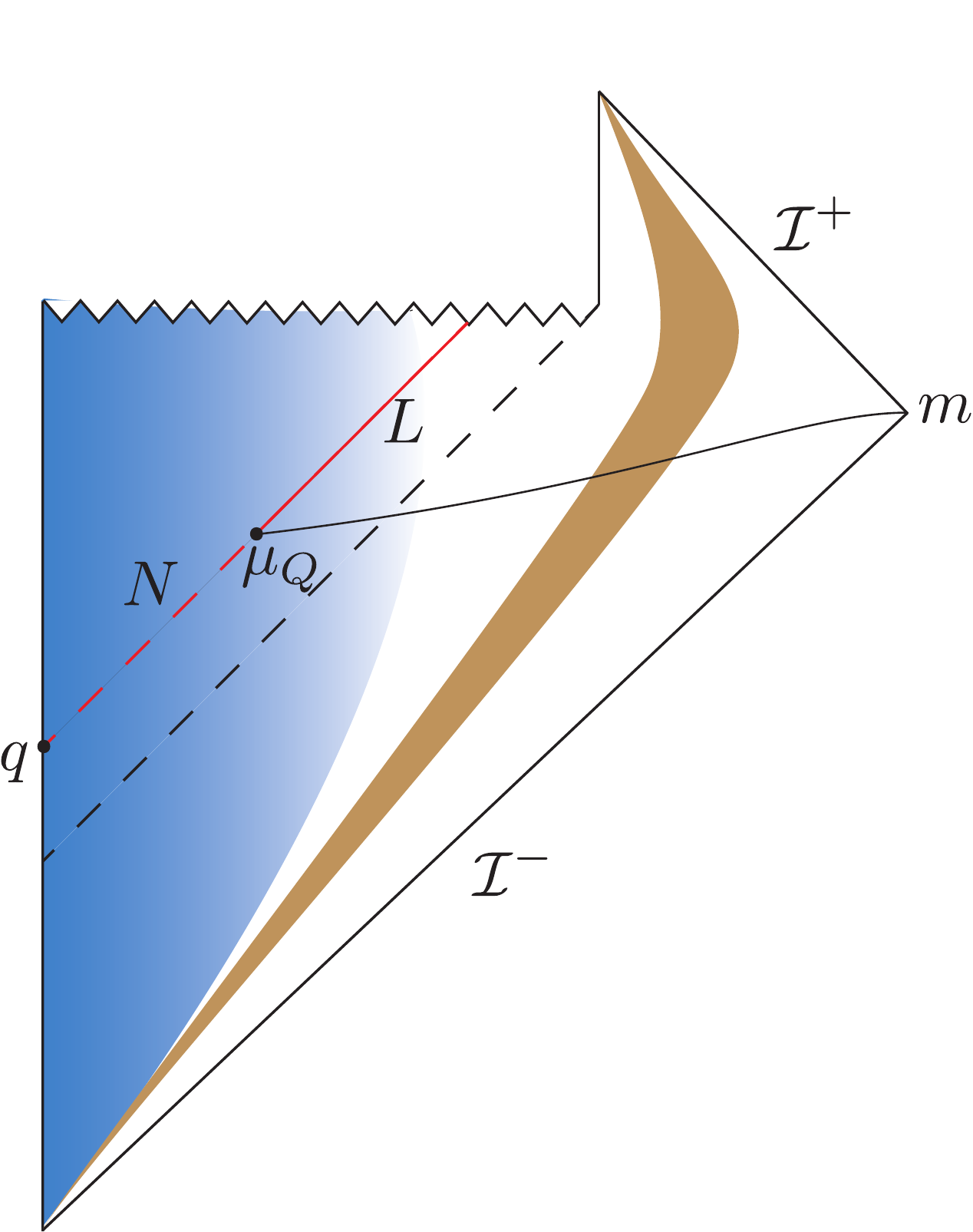}
  %}}%
    %\qquad
   % \subfloat[label 2]{{\includegraphics[width=.425\textwidth]{AdSQPI}}}%
    \caption{The Quantum Penrose Inequality bounds the mass $m$ at infinity in terms of the generalized entropy $S_{\rm gen}$ of any marginally quantum trapped surface $\mu_Q$, evaluated on its future-outgoing lightsheet $L$ (red line). The generalized entropy with respect to the exterior (black slice) can be dominated by distant soft particles (brown) and so does not yield a viable lower bound on the mass.}%
    \label{fig-slices}%
\end{figure}

Finally, we specify on which achronal hypersurface the generalized entropy should be computed. Importantly, this {\em cannot\/} be chosen to be the entire exterior of $\mu_Q$ (the black slice in Fig.~\ref{fig-slices}). This is because an unbounded amount of matter entropy can be present far from the black hole at arbitrarily small cost in energy, such as a dilute gas of photons of large wavelength. 

Moreover, an isolated distant system cannot have net negative energy. In order to deal with the counterexample to the classical Penrose inequality, the QPI will need to have sensitivity only to matter that can be close to the black hole. We will go even further and worry only about matter that actually enters the black hole.

Thus we propose that the generalized entropy should be evaluated on a surface that is entirely contained inside the black hole and has no other boundary with nonzero area. Such a surface may end on the singularity, or at the future endpoints of the horizon. For definiteness, and because it appears to yield a particularly tight bound, we will choose the future-directed outgoing lightsheet $L$ of $\mu_Q$ (see Fig.~\ref{fig-slices}).

A lightsheet is a null hypersurface with everywhere nonpositive expansion $\theta$~\cite{CEB1}. Here we consider what might be called a quantum lightsheet: a null hypersurface orthogonal to $\mu_Q$ with nowhere positive quantum expansion, $\Theta_+\leq 0$. Note that $\Theta_+=0$ on $\mu_Q$ by construction. Assuming the Quantum Focussing Conjecture~\cite{BouFis15a}, the quantum expansion will remain nonpositive everywhere on $L$, so the future-directed outgoing null surface is automatically a quantum lightsheet.

Note that $\mu_Q$ must be quantum marginally trapped with respect to $L$ (and not, for example, with respect to the black slice in Fig.~\ref{fig-slices}). A suitable $\mu_Q$ can be found by picking a null hypersurface $N$ that stays inside the black hole. For example, let $N$ be the future light-cone of an event $q$, see Fig.~\ref{fig-slices}. We can consider the quantum expansion associated with the generalized entropy on $N$, above any cut of $N$. Near $q$, this will be dominated by the large positive classical expansion. Near the singularity, the quantum expansion will be large and negative. Hence $\Theta_+=0$ for some cut in between. In many cases, $\Theta_-\leq 0$ on the same cut (e.g., in the spherically symmetric example shown, and for perturbations around it).

Thus, we propose the Quantum Penrose Inequality
\begin{equation}
  m\geq \sqrt{\frac{\hbar S_{\rm gen}[L(\mu_Q)]}{4\pi G}}
  \label{eq-qpi}
\end{equation}
in asymptotically flat spacetimes. The appearance of a lightsheet suggests an interesting connection with holographic entropy bounds~\cite{CEB1,RMP}, which relate matter entropy to the area of surfaces. However, Eq.~(\ref{eq-qpi}) is different, in that it relates generalized entropy in strongly gravitating regions to the total energy of the spacetime.

\subsection{Preliminary Evidence}
\label{sec-tests}

To gain some intuition, let us test the QPI in the setting of perturbative quantum states on the maximally extended Schwarzschild background. We begin by dismissing the Hartle-Hawking state, in which the black hole is in equilibrium with a thermal bath. The QPI is trivially satisfied, since the mass of the thermal bath diverges.

\begin{figure}[]
\includegraphics[width=.4\textwidth]{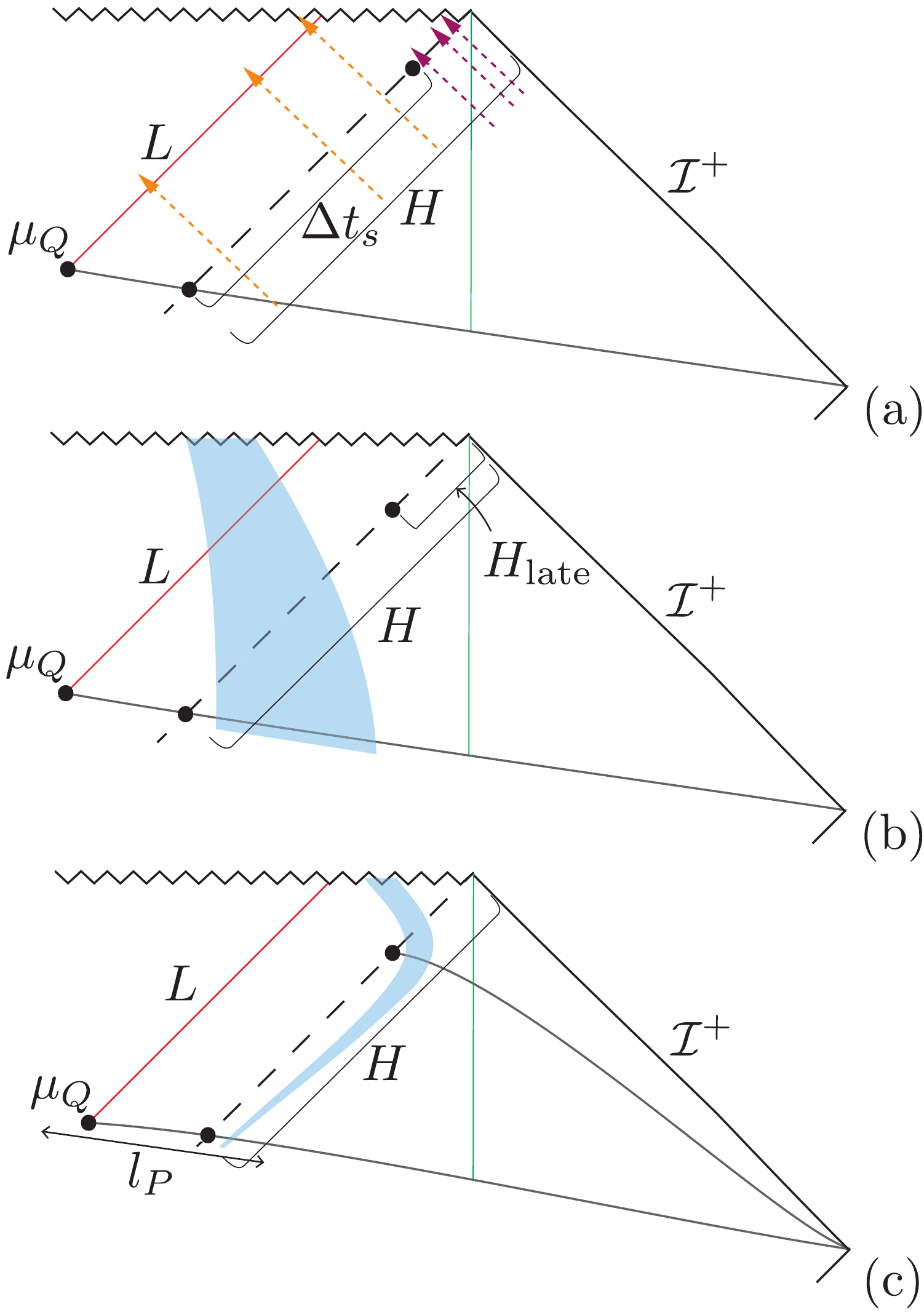}
\caption{Tests of the QPI. (a) An evaporating black hole has a deficit of infalling energy and entropy on the event horizon $H$ (dashed arrows), compared to the Hartle-Hawking state. $S_{\rm gen}[L(\mu_Q)]$ is thus lower than $A[\mu_Q]/4G\hbar$, but only by $\log (R/l_{P})$. Most of the negative entropy fails to reach $L$ (purple arrows). The QPI is nearly saturated. (b) Perturbative state crossing $L$; the QPI is upheld by the GSL. (c) A state with negative energy entering the black hole at sufficiently late times could violate the QPI for $\mu_Q$, since it fails to register on $L$. But this state is not semiclassical near $\mu_Q$.}
\label{fig-3in1}
\end{figure}

A more interesting situation obtains if we turn off the radiation coming in from the right past null infinity ${\cal I}^-$ (Fig.~\ref{fig-3in1}a). Then the black hole will evaporate on the right. Backreaction can be implemented perturbatively near the original bifurcation surface. There will still be a quantum extremal surface and other marginally trapped surfaces. These will lie just inside the (right) event horizon $H$~\cite{Wal13} and will have area smaller than that of $H$ by about one Planck area~\cite{BouEng15c}. Thus the lightsheet $L$ will end at the singularity. We take $\mu_Q$ to be the quantum extremal surface, which will satisfy $16\pi G^2 m^2-A[\mu_Q]\sim O(1)>0$~\cite{BigQPI}. Later quantum marginally trapped surfaces will have smaller area and so are less likely to exhibit a violation of the QPI.
The generalized entropy on $L$ can be estimated by comparing the state to the Hartle-Hawking state, in which $S_{\rm gen}[L] =A[\mu_Q]/4G\hbar$. The difference is that ingoing Hawking radiation is absent, which reduces the entropy, approximately by the number of ingoing modes that cross $L$. On the horizon $H$, the negative entropy of the missing Hawking quanta has the same magnitude as the black hole area, so a bound on $m$ in terms of $S_{\rm gen}[H]$ would be trivial. But after about one scrambling time $\Delta t_s\sim R\log (R/l_P)$, ingoing null geodesics fail to reach $L$ and instead land on the singularity (see Fig.~\ref{fig-3in1}a). Therefore $S_{\rm gen}[L] - A[\mu_Q]/4G\hbar \sim \log (R/l_P)$, and the QPI is nearly saturated:
\begin{equation}
  \frac{4\pi G}{\hbar} m^2-S_{\rm gen}[L(\mu_Q)]\sim \log \frac{R}{l_P}~.
\end{equation}
The logarithmic gap can be eliminated and the QPI even more closely saturated, by preparing an initial state that behaves approximately as a time-reversed Unruh state for one scrambling time~\cite{BigQPI}. 

Next, consider the case where matter crosses the lightsheet $L$ (Fig.~\ref{fig-3in1}b). As explained above, this is only possible if the matter enters the black hole within approximately a scrambling time after $\mu_Q$. If the state is perturbative, $L$ remains close to the event horizon for most of this time. We thus have $S_{\rm gen}[L]\approx S_{\rm gen}[H]-\Delta S[H_{\rm late}]$, where $\Delta S[H_{\rm late}]$ is the matter entropy that crosses $H$ but not $L$. Here we have chosen a state in which this matter is well separated from and has no mutual information with that which crosses $L$. After about a scrambling time, the black hole will settle down to a Kerr solution, for which the generalized entropy will be given by $S_{\rm gen}[H_{\rm late}]-\Delta S[H_{\rm late}]=A_{\rm late}/4G\hbar\leq (4\pi G/\hbar) m^2$.  By the GSL for event horizons, $S_{\rm gen}[H]\leq S_{\rm gen}[H_{\rm late}]$. These results together establish that Eq.~(\ref{eq-qpi}) holds.

Notice that this case was quite general. It includes the Boulware-like counterexample to the classical Penrose inequality. We see how the QPI evades the counterexample: the mass at infinity is too low for Eq.~(\ref{eq-piintro}); but by the GSL, the generalized entropy of $L(\mu_Q)$ is less than the area, by enough for Eq.~(\ref{eq-qpi}) to hold.

On the other hand, if matter with positive entropy enters the black hole within a scrambling time, then the QPI is stronger than Eq.~(\ref{eq-piintro}). By the GSL, the QPI ``knows'' that the black hole is about to grow, so the lower bound on the mass should be adjusted up.

A more dangerous situation would arise if net negative energy were present outside the black hole but failed to pass through the lightsheet $L$ (Fig.~\ref{fig-3in1}c). The QPI predicts that this cannot happen. A nontrivial check on this prediction can be made by preparing a Boulware-like state more than one scrambling time after $\mu_Q$. This negative energy enters the black hole but misses $L$. However, evolving back to $\mu_Q$ blueshifts the fields near the horizon by a factor of at least $\log (\Delta t_s/R) \sim R/l_P$. This implies transplanckian energy densities on and near the horizon, on a Cauchy surface that includes $\mu_Q$. The initial state is not under semiclassical control. We regard this failed counterexample as a nontrivial check on the QPI.

\paragraph*{Acknowledgments}
We thank R.~Emparan, N.~Engelhardt, G.~Horowitz, D.~Marolf,  R.~Myers, J.~Sorce and A.~Wall for discussions. This work was supported in part by the Berkeley Center for Theoretical Physics; by the Department of Energy, Office of Science, Office of High Energy Physics under QuantISED Award DE-SC0019380 and contract DE-AC02-05CH11231; and by the National Science Foundation under grant PHY1820912. MT acknowledges financial support coming from the innovation program under ERC Advanced Grant GravBHs-692951.

\bibliographystyle{utcaps}
\bibliography{all}

\end{document}